\documentclass[3p, twocolumn, sort&compress]{elsarticle}

\usepackage{graphicx}% Include figure files
\usepackage{dcolumn}% Align table columns on decimal point
\usepackage{bm}% bold mathmwe
\usepackage{color}% This is just used temporarily, for notes to myself
\usepackage{subfigure}% Allow for grouped figures.
\usepackage{multirow}% Support for tables
\usepackage{amssymb}
\usepackage{hyperref}
\usepackage{eucal}
\usepackage{lineno}
\newcommand{\iso}[2]{$^{#1}$#2}
\newcolumntype{d}[1]{D{.}{.}{#1} }

\begin{document}
\begin{frontmatter}

\title{Design and demonstration of a quasi-monoenergetic neutron source}
\author[UCB,LLNL]{T.H.~Joshi\corref{cor1}}
\ead{thjoshi@berkeley.edu}

\author[LLNL]{S.~Sangiorgio}
\author[LLNL]{V.~Mozin}
\author[UCB,LLNL]{E.B.~Norman}
\author[LLNL]{P.~Sorensen}

\author[PSU,LLNL]{M.~Foxe\fnref{add}}

\author[LLNL]{G.~Bench}
\author[LLNL]{A.~Bernstein}

\cortext[cor1]{Corresponding author}
\fntext[add]{Current address: Pacific Northwest National Laboratory, USA}

\address[UCB]{Department of Nuclear Engineering, University of California, Berkeley, CA 94720, USA}
\address[LLNL]{Lawrence Livermore National Laboratory, Livermore, CA 94550, USA}
\address[PSU]{Department of Mechanical and Nuclear Engineering, The Pennsylvania State University, University Park, PA 16802, USA}

\begin{abstract}

The design of a neutron source capable of producing 24 and 70 keV neutron beams with narrow energy spread is presented.  The source exploits near-threshold kinematics of the \iso{7}{Li}(p,n)\iso{7}{Be} reaction while taking advantage of the interference `notches' found in the scattering cross-sections of iron.  The design was implemented and characterized at the Center for Accelerator Mass Spectrometry at Lawrence Livermore National Laboratory.  Alternative filters such as vanadium and manganese are also explored and the possibility of studying the response of different materials to low-energy nuclear recoils using the resultant neutron beams is discussed.

\end{abstract}

\begin{keyword}
Neutron source \sep Neutron filter \sep \iso{7}{Li}(p,n)\iso{7}{Be} reaction \sep Nuclear recoil

\end{keyword}

%\pacs{95.35.+d, 95.55.Vj, 25.20.Dc}
\end{frontmatter}
\section{Introduction}

Characterizing the response of radiation detector media to low-energy $\mathcal{O}$(keV) recoiling atoms, often referred to in the literature as nuclear recoils, is necessary to accurately understand the sensitivity of radiation detectors to light weakly interacting massive particles (WIMPS) \cite{Gaitskell,Chepel} and coherent elastic neutrino-nucleus scattering (CENNS) \cite{Hagmann,AkimovCNNS,Freedman,Drukier,Barbeau2}.  To produce nuclear recoils of known energy, several different types of experiments have been proposed; the use of monoenergetic neutron sources and tagging the scattered neutron \cite{Manzur,Gastler,PhysRevC.88.035806}, exploiting time of flight and neutron tagging with a pulsed neutron source \cite{Alexander}, end-point measurements using a monoenergetic neutron source \cite{Jones2,PhysRevLett.110.211101}, use of broad spectrum neutron sources and comparison with monte carlo simulations \cite{Sorensen}, and tagged resonant photo-nuclear scatter \cite{Joshi}.  With the exception of the proposal to use resonant photo-nuclear scatter, these experimental designs have all been employed, however successful characterization of sub-keV nuclear recoils has been limited to several results in germanium \cite{Barbeau2,Jones,Jones2}.  A quasi-monoenergetic $\mathcal{O}$(10 keV) neutron source that can be easily constructed at small proton accelerators would enable further characterization of low-energy nuclear recoils in candidate detector materials.  More generally, such a source would be useful for characterizing the response of detector materials to $\mathcal{O}$(10 keV) neutrons.  

In this article we present the design of a neutron source capable of producing such a beam.  The design employs the near-threshold kinematics of the \iso{7}{Li}(p,n)\iso{7}{Be} reaction to target resonance interference notches present in the neutron scattering cross-section of certain isotopes.  The use of resonance interference notches as neutron filters, only transmitting neutrons within a narrow energy range, has been successfully demonstrated for many years using nuclear reactors \cite{Barbeau,RussianFilter}, however the availability of research reactors instrumented and available for this type of work is limited.  Using a nuclear reaction as the source of neutrons allows production of neutron beams with narrow energy spread at proton accelerator beam-lines capable of producing 2 MeV beams.  

A prototype neutron source was constructed at the target station of the microprobe beam line at the Center for Accelerator Mass Spectrometry (CAMS) at Lawrence Livermore National Laboratory (LLNL) \cite{uprobe}.  In Sec.~\ref{sec:nearthreshold} we discuss the characteristics of near-threshold \iso{7}{Li}(p,n)\iso{7}{Be}.  In Sec.~\ref{sec:filtering} we discuss the use of interference notches in iron, vanadium, or manganese as neutron filters.  The results from characterization of the neutron source using an iron filter  are described in Sec.~\ref{sec:Validation} and a discussion of possible low-energy nuclear recoils measurements that may be performed with such a neutron source is included in Sec.~\ref{sec:Target}.

\section{Near-threshold \iso{7}{Li}(p,n)\iso{7}{Be}}
\label{sec:nearthreshold}

\begin{figure}[t]
\begin{center}
\includegraphics[angle=0,width=0.48\textwidth]{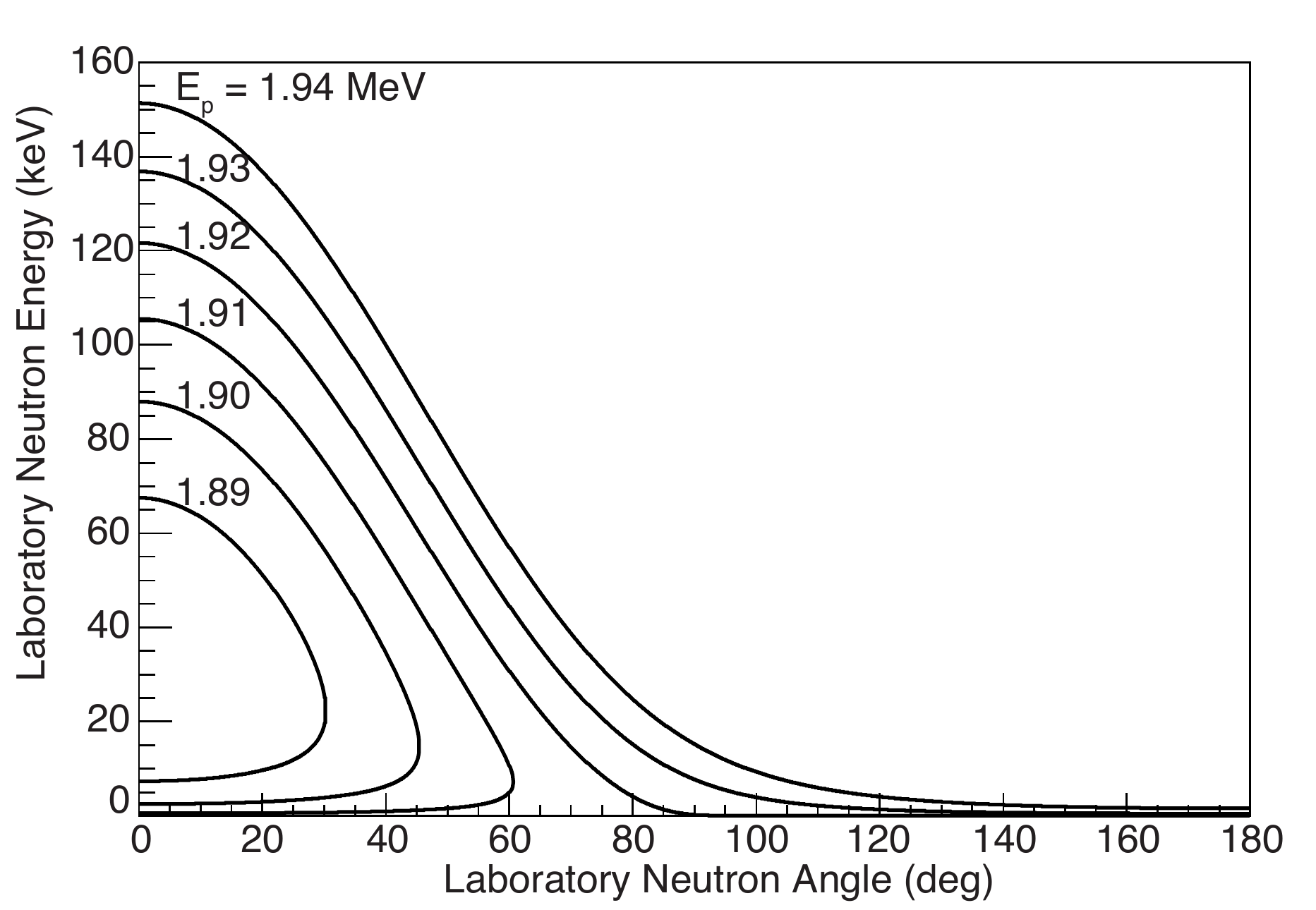}
\caption{Proton energy contours for the \iso{7}{Li}(p,n)\iso{7}{Be} reaction near threshold.}
\label{fig:EnergyContours}
\end{center}
\end{figure}

The \iso{7}{Li}(p,n)\iso{7}{Be} reaction has been extensively studied and used as an accelerator based neutron source thanks to the low Q-value (1.88 MeV) \cite{Drosg}. In the near-threshold regime of the \iso{7}{Li}(p,n)\iso{7}{Be} reaction, the incident energy of the proton beam ($E_{p}$) establishes a kinematically constrained maximum neutron energy that varies with polar angle ($\varphi$) with respect to the incident proton beam.  This behavior is evident in the proton energy contours shown in Fig.~\ref{fig:EnergyContours}.  Though solution of the non-relativistic kinematics equations to understand kinematic constraints of neutron production is straight-forward, calculation of the differential neutron yield is non-trivial.  In this study we employ the prescription given in \cite{Lee} for calculation of near-threshold differential neutron yield for protons traversing a Li-loaded target. This methodology was validated in \cite{Tanaka}.  We make the following reasonable assumptions throughout the article: proton energy loss is constant within the thin targets considered, the incident proton beam is mono-energetic, and target composition is uniform.

\begin{figure*}[]
\begin{center}
\includegraphics[angle=0,width=0.90\textwidth]{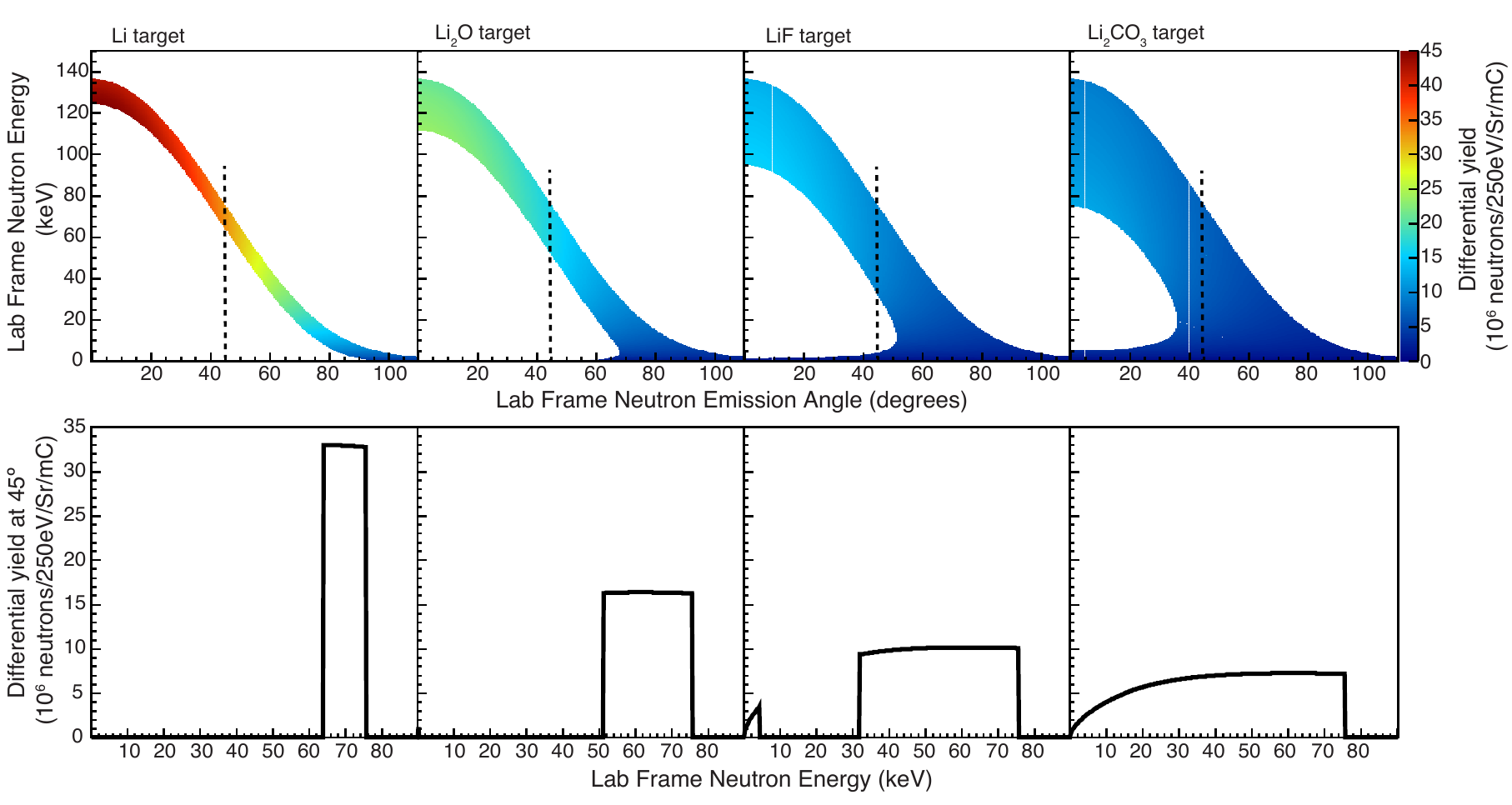}
\caption{(Color online) Comparison of differential neutron yields for different lithium-loaded targets (Proton energy $E_p=1.930$ MeV).  (top) Calculated differential neutron yield in neutrons/mC/250 eV/steradian and (bottom) ideally collimated neutron spectra at $45^{\circ}$ for different thin lithium loaded targets (left to right); 53 $\mu \mbox{g/cm}^2$ metallic lithium, 115 $\mu \mbox{g/cm}^2$ lithium oxide, 199 $\mu \mbox{g/cm}^2$ lithium fluoride, and 285 $\mu \mbox{g/cm}^2$ lithium carbonate.  Areal density selected to keep areal number density of lithium constant.   }
\label{fig:DiffYield}
\end{center}
\end{figure*}

Using this prescription we are able to calculate the expected differential neutron yield for any combination of proton beam energy, lithium-loaded target composition, and target thickness. Figure~\ref{fig:DiffYield} illustrates thin target behavior for 53 $\mu\mbox{g/cm}^2$ metallic lithium, 115 $\mu\mbox{g/cm}^2$ lithium oxide, 199 $\mu\mbox{g/cm}^2$ lithium fluoride, 285 $\mu \mbox{g/cm}^2$ lithium carbonate targets computed in 0.250 keV and $0.5^{\circ}$ intervals with $E_p=1.930$ MeV.  The areal densities were selected such that lithium areal number density is the same for the four example targets. Lithium carbonate, though not a traditional target material, is selected because, as discussed in Sec. \ref{sec:Validation}, the metallic lithium target used to characterize the neutron source was inadvertently mishandled, resulting in a composition of lithium carbonate.  Integrating the differential neutron yield over discrete angles allows comparison of an ideally collimated source with different target characteristics.  Fig.~\ref{fig:DiffYield} compares resultant neutron spectra from these thin targets when collimated at $45^{\circ}$.  Thin targets have several benefits for production of highly tuned neutron sources.  As illustrated in Fig.~\ref{fig:DiffYield}, well collimated thin lithium targets may be used to produce neutron beams with small energy spreads by varying collimation angle and/or proton beam energy.  As a result, thin targets allow kinematic selection of neutron energies and avoid production of extraneous neutrons (those not produced in the desired energy and angular range), thus limiting the experimental backgrounds associated with neutrons (e.g. elastic and inelastic scatter of neutrons and capture gammas).  Additionally, the 478 keV gamma yield from inelastic proton scatter within the lithium-loaded target, \iso{7}{Li}(p,p')\iso{7}{Li}, is significantly reduced when using thin targets.   

\begin{figure}[!h]
\begin{center}
\includegraphics[angle=0,width=0.45\textwidth]{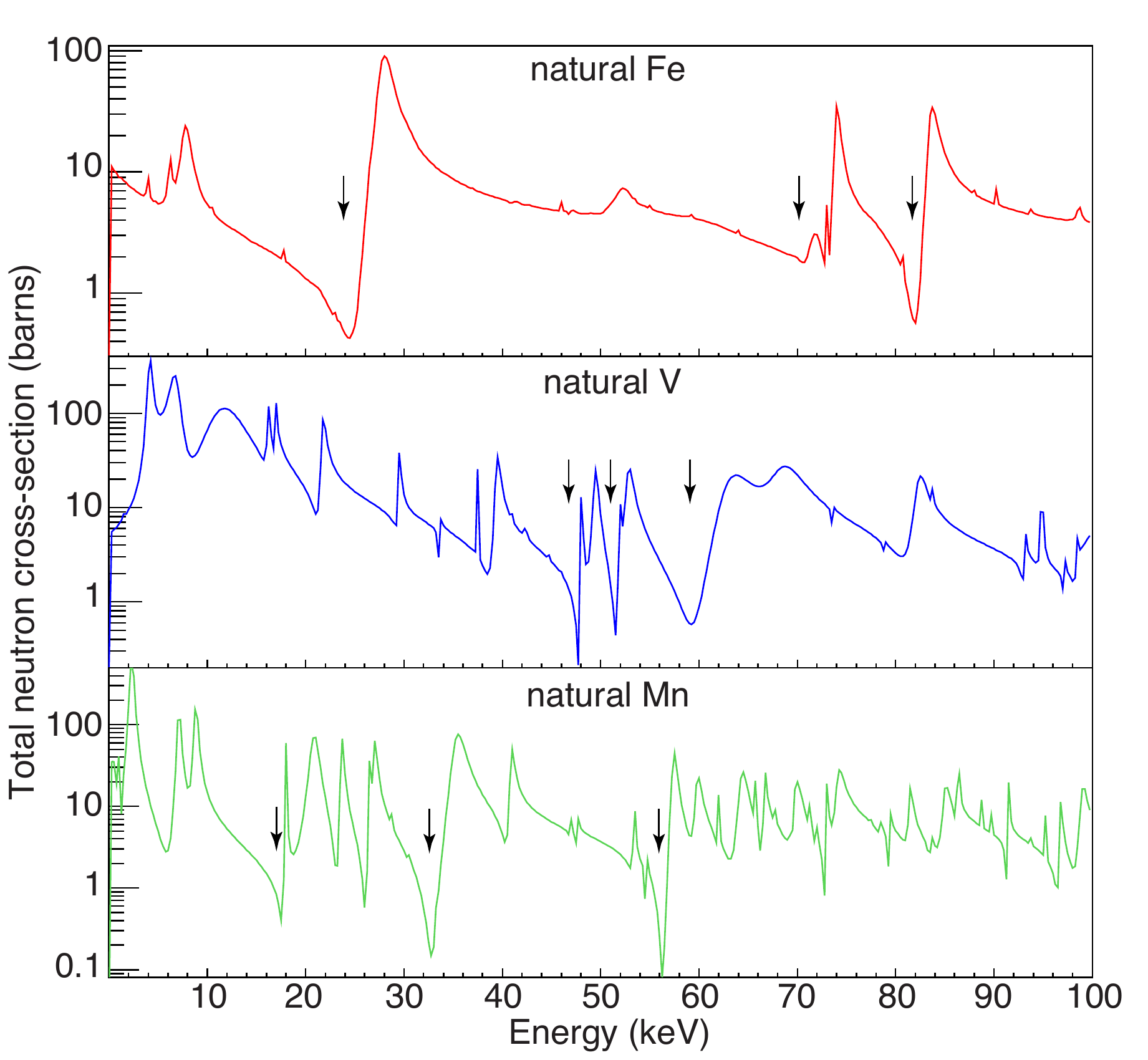}
\caption{(Color online) Illustration of interference notches in the cross-section of Fe, V, and Mn.  ENDF VII total neutron cross-sections for natural compositions of (top to bottom) iron, vanadium, and manganese. Arrows indicate the interference notches that may be used as neutron filters with the near-threshold neutron source. }
\label{fig:XSdata}
\end{center}
\end{figure}

For these reasons very thin targets may be quite attractive for some applications, however production, handling, and lifetime of very thin metallic Li targets pose experimental challenges.  Very thin targets of lithium oxide or lithium fluoride may be used to ease these concerns, but come at the sacrifice of total neutron rate and increased target stopping power (for equivalent lithium areal number density) which broadens the energy of a collimated beam (Fig.~\ref{fig:DiffYield}).  It should also be noted that the neutron energy in sources of this type are entirely defined by reaction kinematics and, as a result, are very sensitive to uncertainties in proton energy and angular alignment.

\section{Filtered neutron beams}
\label{sec:filtering}

One approach to utilize the benefits of near-threshold kinematics for production of beams with narrow energy spread ($\sim$10\% FWHM), while minimizing sensitivity to uncertainties in proton beam energy and angular location, is the exploitation of narrow resonance interference notches in the neutron scattering cross-section of some isotopes.  Interference notches selectively transmit neutrons of a particular energy (Fig.~\ref{fig:XSdata}), allowing some materials to serve as a neutron filter.  A material endowed with an interference notch at a desirable energy may be used as a neutron filter in combination with a collimated \iso{7}{Li}(p,n)\iso{7}{Be} source.  Placement of the filter within the collimator aperture and tuning the kinematics of a near-threshold \iso{7}{Li}(p,n)\iso{7}{Be} source to target the notch effectively produces a neutron beam with narrow energy spread.  The width of the energy spread is dependent upon the properties of the interference notch and the thickness of the filter.  The presence of the filter within the collimator also effectively attenuates the 478 keV gammas produced via inelastic scatter in the target, resulting in a quasi-monoenergetic neutron beam with limited gamma contamination.  

\begin{figure*}[!t]
\begin{center}
\includegraphics[angle=0,width=0.98\textwidth]{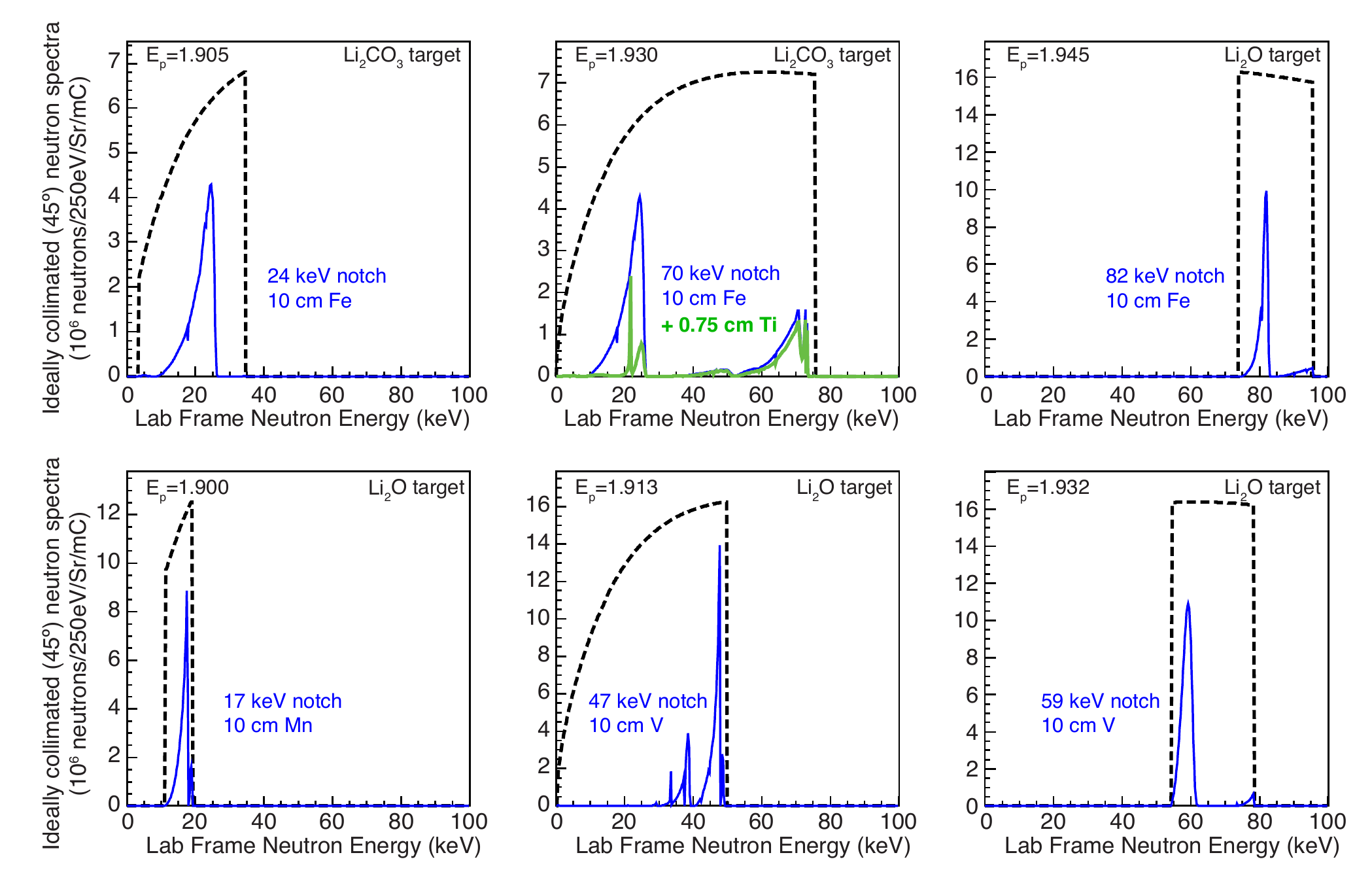}
\caption{(Color online) Calculated unfiltered and filtered neutron spectra targeting interference notches.  Resultant neutron spectra from ideally collimated unfiltered (dashed) and filtered (solid) neutron sources collimated at $45^{\circ}$.  The top left and center spectra use the differential yield of lithium carbonate (285 $\mu\mbox{g/cm}^2$).  Top right and all bottom spectra use the differential yield of lithium oxide (115 $\mu\mbox{g/cm}^2$).  Proton beam energy ($E_p$) used to calculate the differential yields is indicated on each figure.  The length and material of each filter is listed on each figure. }
\label{fig:FilterComparison}
\end{center}
\end{figure*}

The sharp maximum neutron energy dictated by reaction kinematics can be used to target specific interference notches.  Depending on the presence of lower-energy notches within the filter cross-section, and the thickness of the Li-loaded target, the resulting neutron beam may sometimes be composed of more than one spectral components.  To avoid the situation where lower energy notches are filled when targeting higher energies, thin lithium loaded targets may be employed.  Alternatively, an additional material may sometimes be identified that effectively out-scatters the lower energy component while allowing some transmission of the higher energy neutrons, and thus be used as a pre-filter.

The 24 keV notch in iron has been characterized for production of neutron beams at nuclear reactors \cite{Barbeau}. The 24, 70 and 82 keV notches (Fig.~\ref{fig:XSdata}) in natural iron may be targeted using the approach described here. If targeting the 70 or 82 keV notches with a thick Li-loaded target, a titanium filter may be used in combination with the iron to effectively out-scatter the 24 keV neutrons.  Figure~\ref{fig:FilterComparison} illustrates the ideally collimated neutron spectra (before and after filtering) when targeting these candidate notches in iron.  A lithium carbonate differential neutron yield is used in Fig.~\ref{fig:FilterComparison} when illustrating the 24 and 70 keV notches because it was the configuration used to experimentally characterize the neutron source as discussed in Sec.~\ref{sec:Validation}. Targeting of the 82 keV notch is illustrated with a lithium oxide target.  

While iron is an effective neutron filter with several notches, the many naturally occurring isotopes with competing cross-sections limit its performance. An enriched \iso{56}{Fe} filter would perform significantly better than one made with natural iron.  Despite this drawback natural iron was selected for experimental demonstration of this work due to availability.  Several other materials, such as vanadium and manganese, are also endowed with interference notches that may be targeted with the neutron source described and are both composed of single naturally occurring isotopes.  Fig.~\ref{fig:FilterComparison} shows several examples of configurations where these filters may be employed to provide narrow neutron beams with a lithium oxide target.

\section{Experimental demonstration with an iron filter}
\label{sec:Validation}

\begin{figure}[!h]
\begin{center}
\includegraphics[angle=0,width=0.48\textwidth]{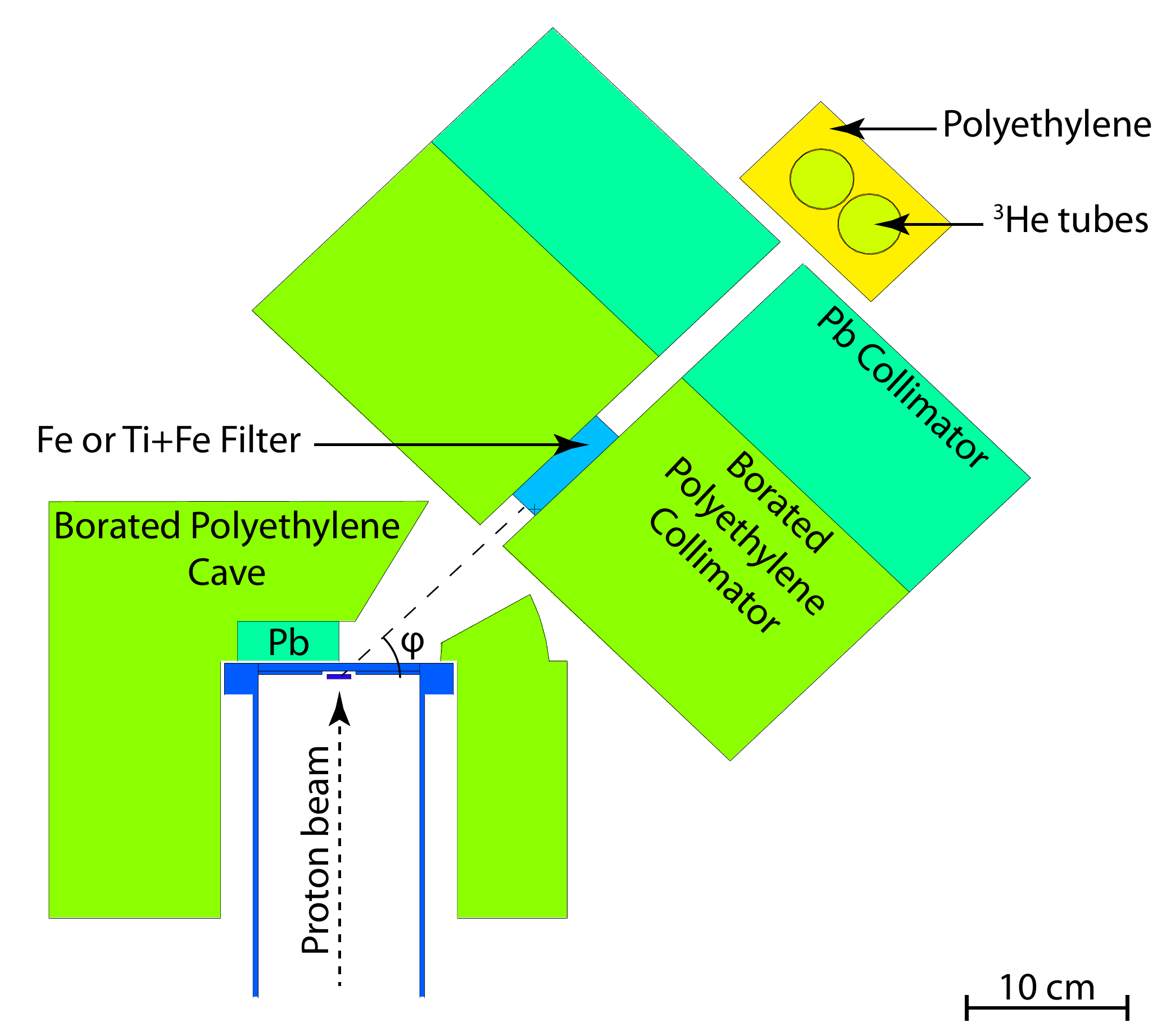}
\caption{(Color online) Schematic of experimental arrangement.  Schematic of the beam on target, borated polyethylene and lead shielding, and counting arrangement (moderated \iso{3}{He} tubes) for validation of the 24 and 70 keV neutron beams. }
\label{fig:Schematic}
\end{center}
\end{figure}

While the performance of the 24 keV notch in iron was well characterized at a reactor facility \cite{Barbeau}, the use of an iron filter in combination with a near-threshold \iso{7}{Li}(p,n)\iso{7}{Be} source has not been demonstrated.  To validate this arrangement several measurements were performed with the prototype neutron source that was built at LLNL.  The experimental setup, illustrated in Fig.~\ref{fig:Schematic}, consisted of an electrically isolated flange at the target chamber of the 1.7 MV National Electrostatic Corporation 5SDH-2 tandem accelerator where the lithium target was held.  The metallic lithium target, 53 $\mu \mbox{g/cm}^2$, evaporated on a 3 mm thick tantalum backing, was inadvertently exposed to small amounts of air during installation.  Rutherford backscatter (RBS) analysis of the target indicates the composition to be lithium carbonate (Li$_{2}$CO$_{3}$) with areal density of $285\pm57$ $\mu \mbox{g/cm}^2$.  Replacement targets were not available so this target was used.  

An Ortec 439 current integrator was connected to the isolated target-holding flange for current integration during measurements, and an Ortec 872 counter/timer was used to count the output of the 439 module. Beam currents were typically 600--700 nA.  A borated polyethylene cave was constructed around the lithium target.  A small 2 cm high path was cut into the shielding from $30^{\circ}-60^{\circ}$ with respect to the proton beam through which neutrons emitted in this direction could pass unimpeded (Fig.~\ref{fig:Schematic}).  One inch of lead shielding was placed within the borated polyethylene cave on the face of the target-holding flange to attenuate the flux of forward directed inelastic scatter gammas produced in the lithium target.  The 2 cm high path was also present in the lead.  An alignment post for the accelerator was placed at $0^{\circ}$ and prevented the experimental setup from exploring shallower angles.  A rotatable table, aligned to the location of the lithium target and outside of the shielding cave, held the borated polyethylene collimator (39.5 x 24 x 15.25 cm) with a 2 cm square beam path.  A 2 x 2 x 7 cm iron filter was placed within the aperture in the borated polyethylene collimator. The collimator was backed by 10.15 cm of lead to attenuate the capture gammas produced in the borated polyethylene shielding. The 2 cm square beam path was also present in the lead shielding.  The rotatable table was able to orient the collimator anywhere from $33^{\circ}-55^{\circ}$ in increments of $\pm0.5^{\circ}$, allowing detectors to be shifted to larger angles where neutron energies are not large enough to pass through the filter due to the reaction kinematics.  All measurements described in this work were obtained with the table oriented at $45^{\circ}$.

The microprobe located at CAMS \cite{uprobe} was used to produce the proton beam.  The energy spread in the proton beam was estimated to be $\pm1$ keV.  The energy of the proton beam was initially calibrated by measuring the threshold for the \iso{7}{Li}(p,n)\iso{7}{Be} reaction using two \iso{3}{He} tubes placed within a  17 x 17 x 7 cm moderator directly behind the Li-target holding flange at $0^{\circ}$, with all shielding and collimator components removed. The \iso{3}{He} tubes, Saint-Gobain model 15He3-608-38SHV, were held at 1300 V and signals read out using Ortec 142PC preamps feeding into Ortec 485 amplifiers and LeCroy 821 discriminators set to accept events within the \iso{3}{He} capture peak and shoulders.  Counts registered by the discriminator were tallied using an Ortec 776 timer/counter.  

\begin{figure*}[t]
\begin{center}
\includegraphics[angle=0,width=0.98\textwidth]{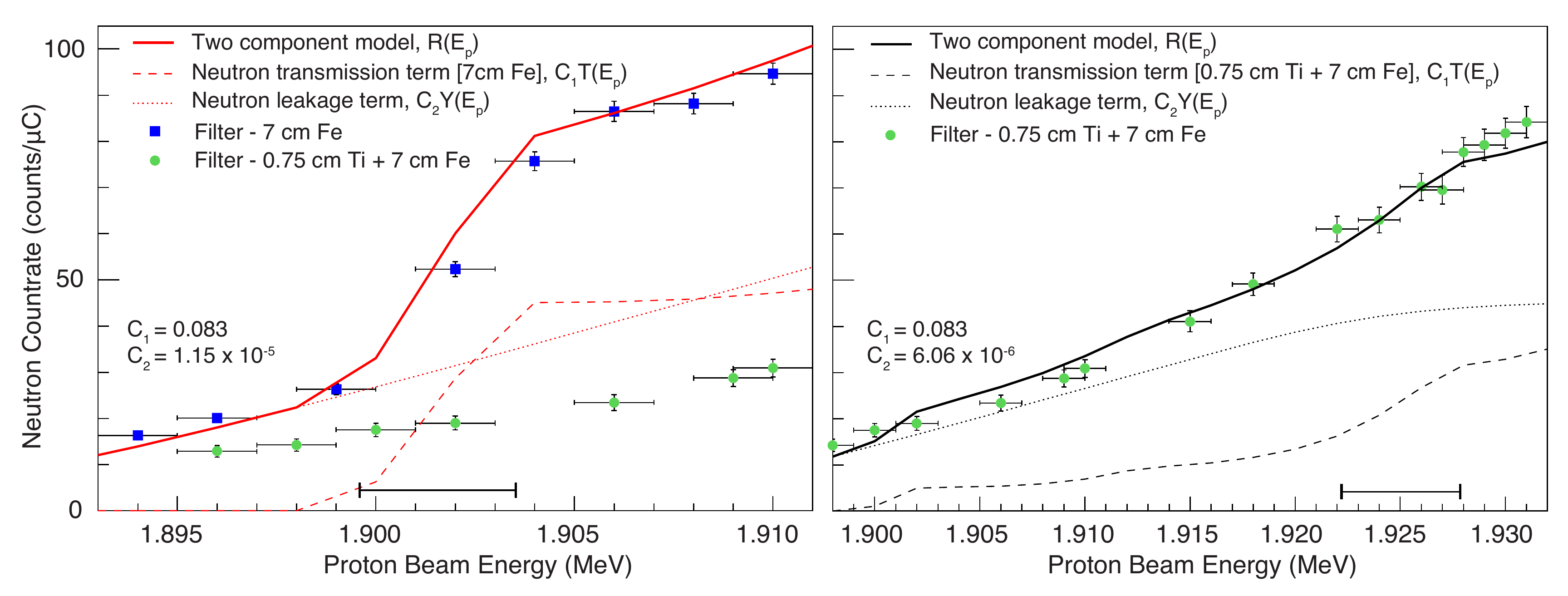}
\caption{(Color online) Experimental demonstration of the kinematic turn on of 24 keV (left) and 70 keV (right) neutrons.  The normalized neutron count rate, using a 7 cm Fe filter (squares) and an additional 0.75 cm Ti pre-filter (circles), was measured at different proton beam energies with \iso{3}{He} tubes placed behind the collimator at $45^{\circ}$ (Fig.~\ref{fig:Schematic}). A model with two free parameters (solid), discussed in the text, matches the data well.  The horizontal bar near the x-axis indicates the region of the 24 and 70 keV kinematic turn on.  A term proportional to the rate of neutrons transmitted through the filter (dashed) ($C_1T$) illustrates the kinematic turn on of 24 and 70 keV neutrons.  A term proportional to the total neutron yield (dotted) ($C_2Y$) accounts for moderated neutrons leaking out of the collimator and shielding.  The values of the free parameters, $C_1$ and $C_2$ are listed on the figure.  Inclusion of a 0.75 cm Ti pre-filter effectively out-scatters 24 keV neutrons removing the kinematic turn on (left). This Ti pre-filter shifts the scattering profile within the collimator, decreasing the magnitude of neutron leakage. The pre-filter was used to isolate contribution of the 70 keV turn on (right). }
\label{fig:Validation}
\end{center}
\end{figure*}

With the shielding and collimator assembled, the performance of the filtered neutron source was validated using the 24 keV transmission notch in iron.  The moderated \iso{3}{He} tubes were placed behind the collimator (Fig.~\ref{fig:Schematic}), aligned to $45^{\circ}$.  Proton beam energy was incrementally increased to observe the kinematic turn-on for 24 keV neutrons.  Figure~\ref{fig:Validation} shows normalized counts (counts/$\mu$C) as proton beam energy is increased.  The kinematic turn-on of the 24 keV notch is clearly visible.  The width of the observed increase in neutron counting rate resultant from 24 keV neutron transmission matches expectations based on the $2^{\circ}$ width of the collimator and the width of the 24 keV notch.  The gradual increase of neutron count-rate before and after the kinematic turn on is attributed to moderated neutrons leaking out of the borated polyethylene collimator and shielding.  The sensitivity of the \iso{3}{He} tubes is larger for the moderated neutrons leaking through the collimator than those transmitted through the filter.  

A two component model of the form $R(E_p)=C_1T(E_p)+C_2Y(E_p)$ was used to describe the data in Fig.~\ref{fig:Validation} (left). R is the neutron count rate, T is the numerically calculated rate of neutrons directed through the collimator and transmitted by the filter at beam energy $E_p$, and Y is the numerically calculated total neutron yield of the source at beam energy $E_p$.  $C_1$ and $C_2$ are free efficiency parameters.  The term $C_1T$ represents the count rate resultant from neutrons transmitted through the collimator and filter.  The term $C_2Y$ describes detected thermal leakage as proportional to the total neutron yield.  The agreement between the experimental data and the two component model indicates the neutron source, collimator, and filter perform as expected, and that moderated neutrons are leaking out of the collimator and shielding.  The rate of moderated neutrons leaking through the shielding is low (\iso{3}{He} tubes are very sensitive to moderated neutrons).  While the escape of some moderated neutrons are unlikely to cause significant backgrounds in experiments that utilize this filtered neutron source, future improvements to the apparatus will include an increase in shielding to reduce this neutron leakage.  

As an additional validation exercise, 0.75 cm of Ti was added to the 7 cm Fe filter and the measurement repeated.  As previously described titanium is very effective at scattering 24 keV neutrons (Fig.~\ref{fig:FilterComparison}) and its inclusion as a pre-filter should remove the feature attributed to the kinematic turn-on of the 24 keV notch. The result of this measurement is also shown in Fig.~\ref{fig:Validation} (left), further confirming that the neutron source and iron filter combination is performing as expected. The titanium filter was placed upstream of the iron filter, shifting the scattering location of neutrons within the collimator upstream and resulting in a reduction of the observed neutron leakage.  Additionally, with the titanium filter installed, the same approach was used to observe the kinematic turn-on of the 70 keV notch Fig.~\ref{fig:Validation} (right). Again the experimental data is well described by the two component model.  The shallow nature of the 70 keV notch, shown in Fig.~\ref{fig:XSdata}, results in the kinematic turn-on being less clear than that of the 24 keV notch.  Above $E_p=1.932$ MeV, neutrons begin to transit the 82 keV notch.  In Fig.~\ref{fig:Validation}, the thermal leakage term, which is proportional to the total neutron yield, plateaus as a result of the thin target.

\section{Applications and discussion}
\label{sec:Target}

The rotatable design of the collimator, though unused in these measurements presented here, may be used to facilitate acquisition of representative background measurements with experiments utilizing this filtered neutron source.  The beam energy ($E_p$) may be tuned such that neutrons effectively transit the filter when the collimator is placed at a shallow angle. A small increase in the collimation angle can then be used to shift the apparatus to a position where the neutron energies that transit the filter are kinematically forbidden. Data acquired at this larger angle will contain the backgrounds present at the shallow angle; associated capture gamma backgrounds from the neutron production, the small flux of lower energy neutrons that may penetrate the collimator and filter, and the 478 keV gammas produced by inelastic proton scatter within the lithium-loaded target.  By fixing $E_p$  and normalizing by integrated proton current on target a background subtraction can be used to isolate signals attributable to neutrons that transit the filter. 

Such measurements may be performed to characterize the response of various materials to quasi-monoenergetic neutrons.  More specifically, the response of materials to low-energy nuclear recoils, which may be produced via elastic neutron scattering, is of interest to the dark matter and CENNS communities.  Table \ref{tab:RecoilEnergy} lists the maximum nuclear recoil energy $E_r(\theta=\pi)=4 E_n*(M m)/(M+m)^2$ from elastic scatter of neutrons with energies of notches illustrated in Fig.~\ref{fig:FilterComparison} on different detector materials.  $E_n$ is the energy of the incident neutron, $M$ is the mass of the target nucleus, and $m$ is the mass of the neutron.  

The accessible nuclear recoil energies are lower than the lowest reported characterization measurement in liquid xenon \cite{Manzur}.  Using such a source to perform ionization and scintillation yield measurements for $\mathcal{O}$(keV) nuclear recoils would provide the necessary information to clarify the sensitivity of dark matter searches using these target materials, allow calculation of the sensitivity of these materials to CENNS of reactor neutrinos, and study the field dependence of recombination following energy deposition.  One such measurement has been demonstrated using this neutron source (70 keV) to measure the ionization yield of 6.7 keV nuclear recoils in liquid argon \cite{Joshi2}.

\begin{table}[t]
\caption{Maximum nuclear recoil energies from filtered neutron beams on Xe, Ar, Ge, and Si.}
\begin{center}
\begin{tabular}{cd{-1}d{-1}d{-1}d{-1}}
\hline \hline
\multicolumn{1}{c}{\emph{Neutron energy}} 
&\multicolumn{4}{c}{\emph{Max recoil energy }(\mbox{keV})} \\
\cline{2-5}
(\mbox{keV})  &  \multicolumn{1}{c}{\emph{Xe}}   &  \multicolumn{1}{c}{\emph{Ar}} & \multicolumn{1}{c}{\emph{Ge}} & \multicolumn{1}{c}{\emph{Si}}  \\
\hline \hline
\multicolumn{1}{c|}{17} & 0.5 & 1.6 & 0.9 & 2.3 \\
\multicolumn{1}{c|}{24} & 0.7 & 2.3 & 1.3 & 3.2 \\
\multicolumn{1}{c|}{47} & 1.4 & 4.5 & 2.5 & 6.3 \\
\multicolumn{1}{c|}{59} & 1.8 & 5.7 & 3.2 & 7.9 \\
\multicolumn{1}{c|}{70} & 2.1 & 6.7 & 3.8 & 9.4 \\
\multicolumn{1}{c|}{82} & 2.5 & 7.9 & 4.4 & 11.0 \\

\hline \hline
\end{tabular}
\end{center}
\label{tab:RecoilEnergy}
\end{table}

\section{Conclusions}

The near-threshold kinematics of the \iso{7}{Li}(p,n)\iso{7}{Be} reaction combined with the neutron transmission properties of materials such as iron, vanadium, and manganese provide the ability to produce neutron beams with narrow energy spreads using a small proton accelerator.  We have designed such a source, and demonstrated production of 24 and 70 keV neutron beams using an iron filter.  This neutron source may be useful for measuring the response of relevant detector materials to $\mathcal{O}$(10 keV) neutrons.  One specific application being the study of detector response to low energy nuclear recoils.  Measurements of this type can provide information about the energy loss mechanisms of low-energy nuclear recoils, the recombination effects when electric fields are present within the targets, and the sensitivity of different detector media.  Additionally, such characterization would enable accurate calculation of the sensitivity of different detector media to CENNS of reactor anti-neutrinos. 

% Acknowledgements 
The authors would like to thank Vincent Meot for production of the Li targets, Sergey Kucheyev for Rutherford Backscatter measurements of the lithium targets, Jason Burke for his help with the experimental effort, and Tom Brown for his assistance at CAMS.  T.H.~Joshi would like to acknowledge the funding of the Lawrence Scholars Program at LLNL and the Department of Homeland Security.  A portion of M. Foxe's research was performed under the Nuclear Forensics Graduate Fellowship Program, which is sponsored by the U.S. Department of Homeland Security, Domestic Nuclear Detection Office and the U.S. Department of Defense, Defense Threat Reduction Agency. We gratefully acknowledge the LDRD program (LDRD 13-FS-005) at LLNL.  This work was performed under the auspices of the U.S. Department of Energy by Lawrence Livermore National Laboratory under contract DE-AC52-07NA27344. LLNL-JRNL-651154%

\bibliographystyle{elsarticle-num}
\bibliography{Joshi_pn_bib}

\end{document}